\documentclass[english]{article}
\usepackage[T1]{fontenc}
\usepackage[latin9]{inputenc}
\usepackage[letterpaper]{geometry}
\usepackage{amssymb}

\makeatletter
\newcommand{\lyxaddress}[1]{
\par {\raggedright #1
\vspace{1.4em}
\noindent\par}
}

\makeatother

\usepackage{babel}

\begin{document}

\title{6-loop anomalous dimension of a single impurity operator from AdS/CFT
and multiple zeta values}

\author{Zoltán Bajnok$^{a}$ and Omar el Deeb$^{b}$}

\maketitle

\lyxaddress{\begin{center}
\emph{$^{a}$Theoretical Physics Research Group of the Hungarian
Academy of Sciences,}\\
\emph{$^{b}$Institute for Theoretical Physics, Roland Eötvös University,
}\\
\emph{H-1117 P\'azm\'any s. 1/A, Budapest, Hungary}
\par\end{center}}

\vspace{-8cm}

\hspace{11cm}ITP-Budapest Report 648

\vspace{8cm}
\begin{abstract}
Anomalous dimension of the simplest nontrivial single impurity operator
in the $\beta=\frac{1}{2}$ deformed theory is determined at six loops
from the AdS/CFT correspondence. Lüscher correction is evaluated at
next-to-next-to-leading order (NNLO) in terms of multiple zeta values.
The result can be simplified into the products of simple zeta functions
and the same form of the correction is expected for the Konishi operator
at six loops, too. 
\end{abstract}

\section{Introduction}

AdS/CFT correspondence \cite{Maldacena:1997re} relates the string
energies on the $AdS_{5}\times S^{5}$ background to the anomalous
dimensions of gauge invariant operators in maximally supersymmetric
four dimensional $SU(N)$ gauge theory. The correspondence is particularly
useful in the large $N$ limit when it can be described by a two dimensional
integrable field theory, see \cite{Arutyunov:2009ga} for a general
introduction and references therein. The advantage of the integrable
two dimensional point of view lies in its non-perturbative nature,
which nevertheless, can be used to calculate perturbative quantities,
too. Indeed, the leading finite size correction of a two particle
state can be described in terms of the asymptotical Bethe ansatz and
the generalized Lüscher formulas \cite{Bajnok:2008bm}, which, when
expanded in the coupling constant, provides the exact perturbative
anomalous dimension of the Konishi operator up to seven loops. The
direct perturbative gauge theory calculations are very cumbersome
and have been evaluated for the Konishi operator \emph{up to 4 loops}
\cite{Fiamberti:2007rj,Fiamberti:2008sh} \emph{only}, where the wrapping
part of the correction matches exactly the Lüscher type finite size
correction \cite{Bajnok:2008bm}. Thus the integrability based {}``string
theory'' techniques provide a way to go beyond the available perturbative
calculations and collect information about higher order results. The
simplest situation in this respect is the calculation of the anomalous
dimension of a single impurity operator. Such operator is trivial
in the $\mathcal{N}=4$ SYM but has non-vanishing anomalous dimension
in orbifold or $\beta$-deformed theories. The leading and next-to-leading
wrapping correction have been calculated in \cite{Bajnok:2008qj}
and in \cite{Bajnok:2009vm} and the aim of the present paper is to
extend the calculation to next-to-next-to-leading order (NNLO). This
order is important in the sense that it is not affected by double
wrapping corrections thus the calculation based on the Lüscher formula
provides and exact answer.

The $\beta$-deformed theory is an $\mathcal{N}=1$ supersymmetric,
exactly marginal deformation of the $\mathcal{N}=4$ SYM and has been
always the testing ground of the AdS/CFT duality. The dispersion relation
of the excitations was calculated exactly in \cite{Mauri:2005pa},
which together with wrapping corrections \cite{Fiamberti:2008sm,Fiamberti:2008sn}
provides the exact anomalous dimension of a single impurity operator
$\mbox{Tr}(XZ^{L})$. On the string theory side this operator corresponds
to a one particle state in the $su(2)$ sector in finite volume. The
leading finite size effect follows from the momentum quantization,
or asymptotical Bethe Ansatz equation \cite{Beisert:2005if}. As the
volume decreases Lüscher/wrapping type correction becomes important
\cite{Bajnok:2008bm,Bajnok:2008qj,Ahn:2010yv} and for an exact description
they have to be summed up. An educated way to take all finite size
correction into account would be to extend the Thermodynamic Bethe
Ansatz of the $\mathcal{N}=4$ theory \cite{Gromov:2009tv,Arutyunov:2009zu,
Bombardelli:2009ns,Arutyunov:2009ur,Gromov:2009bc,Arutyunov:2009ax,Cavaglia:2010nm}
to the $\beta$-deformed case. The first step into this direction
is done: in \cite{Gromov:2010dy} the authors extended the functional
$Y$ system relations, while in \cite{Ahn:2010ws} the scattering
matrix of the theory together with its twisted boundary condition
were identified. 

We are interested in the $\beta$ -deformed theory at a particular
value of $\beta$, namely for the simplest nontrivial $\beta=\frac{1}{2}$
and evaluate the Lüscher correction at NNLO. This value of $\beta$
is interesting as it correspond to a particular orbifold theory, too.
As it was conjectured in \cite{Gunnesson:2009nn} the anomalous dimension
of the $su(2)$ particle ($\mbox{Tr}(XZ^3)$) in the $\beta=\frac{1}{2}$-deformed
theory coincides with the anomalous dimension of the $sl(2)$ particle
($\mbox{Tr}(DZZ)$) of the orbifold theory. A clear description of
the relation can be found in \cite{Arutyunov:2010gu} where a TBA
equation was also proposed to describe the single impurity excitation
of the orbifold theory. The proposed infinite coupled integral equations
in principle exactly describe the energy of the one particle state
for any volume and coupling. However, as it was shown in \cite{Arutyunov:2010gu},
following the work of \cite{Arutyunov:2010gb,Balog:2010xa,Balog:2010vf},
it reduces in the weak coupling limit to the Lüscher correction and
that is what we are going to evaluate now. We will focus on the structure
of the result, especially how the various terms of the perturbation
theory appear in terms of multiple zeta values. 

Multiple zeta values (MZVs) are the generalization of zeta functions:
\[
\zeta(a_{1},a_{2},\dots,a_{n})=\sum_{j_{1}>j_{2}>\dots>j_{n}>0}\frac{1}
{j_{1}^{a_{1}}j_{2}^{a_{2}}\dots\, j_{n}^{a_{n}}}\]
and are trivially related to the values of nested harmonic sums taken
at infinity. They appear at various places of physics and mathematics
including number theory, combinatorics, quantum field theory, statistical
physics. For a recent review see \cite{Bailey:2010aj} and references
therein. What we would like to describe here is how the various MZVs
appear at a given order of the calculation. We will see that at leading
order we have simple zeta function, the next-to-leading order double
MZVs will appear, while the novel result shows the appearance of triple
MZVs at NNLO. Not only the depth of the MZVs increases with the orders
but also their transcendentality degree. There are several nontrivial
relations among the MZVs of a given transcendentality degree and one
can always choose a good basis \cite{Blumlein:2009cf}. Any basis
above transcendentality $8$ contain MZVs which are irreducible in
the sense that cannot be written as products of simple zeta functions.
As the anomalous dimension of the simplest nontrivial operator at
NLO had transcendentality $7$ such irreducible MZVs did not appear
\cite{Bajnok:2009vm}. At the NNLO the transcendentality is $9$ so
we expect the appearance of irreducible MZVs. Let us anticipate our
result. 

We calculate the anomalous dimension of the operator $\mbox{Tr}(XZ)$
in the $\beta=\frac{1}{2}$ deformed theory, or equivalently the anomalous
dimension of the $\mbox{Tr}(DZZ)$ operator in the $\mathbb{Z}_{2}$
orbifold theory. On the string theory side they correspond to the
energy of a one particle state in finite volume $L=1$. It is a very
special state as it has a vanishing rapidity $u=0$, ($p=\pi$), which
is not effected by finite size corrections. In contrast, the energy
is shifted by vacuum polarization effects as \[
E=E_{ABA}+\Delta E\]
The asymptotical Bethe Ansatz energy $E_{ABA}$ is simply the dispersion
relation of a standing particle:\begin{eqnarray*}
E_{ABA} & = & \sqrt{1+16g^{2}\sin^{2}(\frac{p}{2})}=\sqrt{1+16g^{2}}\\
 & = & 1+8g^{2}-32g^{4}+256g^{6}-2560g^{8}+28672g^{10}-344064g^{12}+O\left(g^{13}\right)\end{eqnarray*}
while $\Delta E$ corresponds to the wrapping interactions and has
the expansion \[
\Delta E=\Delta E_{4}g^{4}+\Delta E_{5}g^{10}+\Delta E_{6}g^{12}+\dots\]
In \cite{Bajnok:2008qj} the LO correction was calculated \[
\Delta E_{4}=128(4\zeta(3)-5\zeta(5))\]
The NLO correction turned out to be \cite{Bajnok:2009vm}: \[
\Delta E_{5}=-128(12\zeta(3)^{2}+32\zeta(3)+40\zeta(5)-105\zeta(7))\]
 In the main part of the paper we evaluate the NNLO correction and
obtain 
\begin{equation}
\Delta E_{6}=-128(48\zeta(3)^{2}-592\zeta(5)-24\zeta(3)(8+15\zeta(5))-322\zeta(7)+1701\zeta(9))
\end{equation}
which completes the anomalous dimension of the operator up to six
loops. One can observe that the transcendentality is $9$ and that
no irreducible MZV appeared. This is due to an unexpected cancellation
during the calculation. Let me note also, that the form of the correction
matches exactly with other 6 loop results like the ones in \cite{Velizhanin:2010cm,Arutyunov:2010gu}. 

The rest of the paper is organized as follows: In the next Section
we present the Lüscher correction for a one particle state in the
$\beta=\frac{1}{2}$ deformed theory by recalling the available formulas
from the literature. Section 3 contains the results of the calculation.
As we used a novel method to sum up the various terms we present the
four and five loop results in this fashion and show how we obtained
the result at six loops. We pay particular attention how the various
MZVs appear and describe how they can be simplified. Finally we conclude
in section 4 and explain how our novel results could be used. The
Appendix contains some technical information, especially on how we
summed up nested harmonic sums.

\section{Calculation of the anomalous dimension from AdS/CFT}

As the anomalous dimension of the single impurity operator corresponds
to a standing particle state, we have to calculate the standard Lüscher
correction. It describes how the one particle energy (dispersion relation)
\[
E=E(p=\pi)=\sqrt{1+16g^{2}\sin^{2}(\frac{p}{2})}=\sqrt{1+16g^{2}}=\epsilon\]
is modified due to vacuum polarization effects. The $p=\pi$ momentum
corresponds to vanishing rapidity $u=0$ which is protected by symmetry.
Consequently the leading finite size correction is the energy correction
which originates from virtual particles propagating around the circle
and can be written as\begin{equation}
\Delta E=-\sum_{Q=1}^{\infty}\int\frac{dq}{2\pi}\mbox{sTr}(S_{Q1}^{Q1}(q,0))
e^{-\tilde{\epsilon}_{Q}(q)L}+O(g^{16})\label{eq:Luscher}\end{equation}
Here we sum up for all bound-states of charge $Q$ of the mirror model,
whose momenta are $q$, $S_{Q1}^{Q1}$ describes how they scatter
on the fundamental particle and $\tilde{\epsilon}(q)$ denotes their
mirror energy. We have to expand this expression in NNLO in $g$.
The above form exactly describes the energy correction up to the order
$g^{14}$ only since at the order $g^{16}$ double wrapping effects
will contribute, too. Let us analyze the $g$ dependence of the various
terms.

\subsubsection*{Exponential factor}

The mirror energy has the following parametrization\[
e^{-\tilde{\epsilon}_{Q}(q)}=\frac{z^{-}(q,Q)}{z^{+}(q,Q)}\quad;\qquad z^{\pm}(q,Q)=
\frac{q+iQ}{4g}\left(\sqrt{1+\frac{16g^{2}}{q^{2}+Q^{2}}}\pm1\right)\]

\subsubsection*{Matrix part of the scattering matrix}

The contribution of the scattering matrix can be factored as \[
\mbox{sTr}(S_{Q1}^{Q1}(q,0))=S_{scalar}(q,0)\mbox{sTr}(S_{matrix}^{su(2)}(q,0))^{2}\]
where the super-trace of the matrix part contain the contributions
of all polarizations. The various polarizations of the mirror bound-states
can be labeled in the super-space formalism as $(w_{3}^{j}w_{4}^{Q-j}$,
$w_{3}^{j}w_{4}^{Q-2-j}\theta_{1}\theta_{2}$, $w_{3}^{j}w_{4}^{Q-1-j}\theta_{3}$,
$w_{3}^{j}w_{4}^{Q-1-j}\theta_{4})$ see \cite{Arutyunov:2008zt,Bajnok:2008bm}
for the details. The super-trace in the $\beta$- deformed theory
evaluates in the previous basis as \[
\mbox{sTr}(S_{matrix}^{su(2)}(z,x))=\sum_{j=0}^{Q}SB1_{j}(z,x)+\sum_{j=0}^{Q-2}SB2_{j}(z,x)
+i\sum_{j=0}^{Q-1}SF1_{j}(z,q)-i\sum_{j=0}^{Q-1}SF2_{j}(z,q)\]
The S-matrix elements can be extracted from \cite{Bajnok:2008bm}.
In calculating the corrections for the $su(2)$ representative neither
of the S-matrix contributions depends on $j$ and they read explicitly
as \[
SB1_{j}(z,x)=\frac{z^{+}-x^{+}}{z^{-}-x^{+}}\frac{\tilde{\eta}_{1}}{\eta_{1}}\quad;
\qquad SB2_{j}(z,x)=\frac{z^{+}-x^{-}}{z^{-}-x^{+}}\frac{(1-x^{+}z^{-})}{(1-x^{-}z^{-})}
\frac{x^{-}}{x^{+}}\frac{\tilde{\eta}_{1}}{\eta_{1}}\left(\frac{\tilde{\eta}_{2}}{\eta_{2}}\right)^{2}\]
\[
SF1_{j}(z,x)=\frac{z^{+}-x^{-}}{z^{-}-x^{+}}\frac{\tilde{\eta}_{1}}{\eta_{1}}\frac{\tilde{\eta}_{2}}{\eta_{2}}
\quad;\qquad SF2_{j}(z,x)=\frac{z^{+}-x^{+}}{z^{-}-x^{+}}\frac{(1-x^{+}z^{-})}{(1-x^{-}z^{-})}
\frac{x^{-}}{x^{+}}\frac{\tilde{\eta}_{1}}{\eta_{1}}\frac{\tilde{\eta}_{2}}{\eta_{2}}\]
The appearing string frame factors can be written as $\frac{\tilde{\eta}_{1}}{\eta_{1}}=
\sqrt{\frac{z^{-}}{z^{+}}}$
and $\left(\frac{\tilde{\eta}_{2}}{\eta_{2}}\right)^{2}=\frac{x^{-}}{x^{+}}$.
The $x^{\pm}$ parameters depend on the momentum the usual way \[
x^{\pm}(p)=\frac{\cot\frac{p}{2}\pm i}{4g}\left(1+\sqrt{1+16g^{2}\sin^{2}\frac{p}{2}}\right)\]
which in our case results in \[
x^{\pm}=\pm x=\pm\frac{i}{4g}(1+\epsilon)\quad;\qquad\frac{x^{-}}{x^{+}}=-1\]
With these variables the super-trace of the matrix part takes a particularly
simple form\[
\mbox{sTr}(S_{matrix}^{su(2)}(z,x))=\frac{2x(1+2Qz^{-}(x-z^{+}))}{(x-z^{-})(1+xz^{-})}e^{-\tilde{\epsilon}_{Q}(q)/2}\]
We have checked that the matrix part of the $sl(2)$ representative
of the orbifold model gives the same result. There, the deformation
is such that the fermions do not contribute as the two undeformed
fermionic S matrix elements are the same \cite{Bajnok:2008qj,Lukowski:2009ce}.

\subsubsection*{Scalar part of the scattering matrix}

The scalar part of the scattering matrix of a charge $Q$ bound-state
can be obtained by multiplying the scalar factors of its individual
scattering constituents. The charge $Q$ bound-state composed of elementary
magnons as $z=(z_{1},\dots,z_{Q})$, such that $z^{-}=z_{1}^{-}$
and $z_{Q}^{+}=z^{+}$and the bound-state condition is also satisfied
$z_{i}^{+}=z_{i+1}^{-}$. Thus the full scalar factor as the product
of the elementary scalar factors turns out to be \cite{Bajnok:2008bm}:
\begin{equation}
S_{scalar}(z,x)=\prod_{i=1}^{Q}S^{sl(2)}(z_{i},x)=\prod_{i=1}^{Q}e^{-2i\sigma(z_{i},x)}
\frac{z_{i}^{-}-x^{+}}{z_{i}^{+}-x^{-}}\frac{1-\frac{1}{z_{i}^{+}x^{-}}}{1-
\frac{1}{z_{i}^{-}x^{+}}}\end{equation}
In calculating the Lüscher correction we have to evaluate this expression
when $z$ is in the mirror kinematics ($\vert z^{-}\vert<1,\vert z^{+}\vert>1$).
The analytical continuation has been carefully elaborated in \cite{Arutyunov:2009kf}:

\begin{equation}
S_{scalar}^{-1}(z,x)=\Sigma_{Q,1}^{2}(z,x)S^{su(2)}(z,x)\end{equation}
A particularly good feature of the formula is that both expressions
depend on $z^{\pm}$ only and not on the individual $z_{i}^{\pm}$.
Explicitly the $su(2)$ scalar factors read as: 

\[
S^{su(2)}(z,x)=\frac{(z^{+}-x^{-})(z^{+}-x^{+})}{(z^{-}-x^{+})(z^{-}-x^{-})}
\frac{(1-\frac{1}{z^{+}x^{-}})(1-\frac{1}{z^{+}x^{+}})}{(1-\frac{1}{z^{-}x^{+}})(1-\frac{1}{z^{-}x^{-}})}\]
while for the case $\vert x^{\pm}\vert>1$ following \cite{Arutyunov:2009kf}
we can write \begin{eqnarray*}
-i\log\Sigma_{Q,1}(z,x) & = & \Phi(z^{+},x^{+})-\Phi(z^{+},x^{-})-\Phi(z^{-},x^{+})+\Phi(z^{-},x^{-})\\
 &  & +\frac{1}{2}\left[-\Psi(z^{+},x^{+})+\Psi(z^{+},x^{-})-\Psi(z^{-},x^{+})+\Psi(z^{-},x^{-})\right]\\
 &  & +\frac{1}{2i}\log\left[\frac{(z^{+}-x^{+})(x^{-}-\frac{1}{z^{+}})^{2}}{(z^{+}-x^{-})(x^{-}-
 \frac{1}{z^{-}})(x^{+}-\frac{1}{z^{-}})}\right]\end{eqnarray*}
We have to be careful as the conventions of \cite{Arutyunov:2009kf}
are different from ours. To turn into our conventions one has to replace
$z^{\pm}\to z^{\mp}$ and $x^{\pm}\to x^{\mp}.$ There are integral
representations for $\Phi$ and$\Psi$ as 

\[
\Phi(x_{1},x_{2})=i\oint_{C_{1}}\frac{dw_{1}}{2\pi i}\oint_{C_{1}}\frac{dw_{2}}{2\pi i}
\frac{1}{w_{1}-x_{1}}\frac{1}{w_{2}-x_{2}}\log\frac{\Gamma(1+ig(w_{1}+w_{1}^{-1}-w_{2}-w_{2}^{-1}))}
{\Gamma(1-ig(w_{1}+w_{1}^{-1}-w_{2}-w_{2}^{-1}))}\]

\[
\Psi(x_{1},x_{2})=i\oint_{C_{1}}\frac{dw_{2}}{2\pi i}\frac{1}{w_{2}-x_{2}}\log\frac{
\Gamma(1+ig(x_{1}+x_{1}^{-1}-w_{2}-w_{2}^{-1}))}{\Gamma(1-ig(x_{1}+x_{1}^{-1}-w_{2}-w_{2}^{-1}))}\]
where the integrations are for the unit circle. They are well-defined
provided none of the $x_{i}$ lies on the unit circle. 

As by now we have collected all the necessary formulas we turn to
analyze their weak coupling expansions.

\subsubsection*{Weak coupling expansion}

Our aim is to calculate the weak coupling expansion of $\Delta E$
for $L=1$. In doing so we decompose the integrand of the Lüscher
correction (\ref{eq:Luscher}) \[
\Delta E=-\sum_{Q=1}^{\infty}\int\frac{dq}{2\pi}P(q,Q)\Sigma(q,Q)\]
into a simpler rational part \[
P(q,Q)=\frac{4x^{2}(1+2Qz^{-}(x-z^{+}))^{2}}{((xz^{-})^{2}-1)(x^{2}-(z^{+})^{2})}\left(\frac{z^{-}}{z^{+}}\right)^{2}\]
which contains both the matrix part and the rational part of the scalar
factor, and into the more complicated $\Sigma$ part: \begin{eqnarray*}
i\log\Sigma(q,Q) & = & 2(\Phi(z^{+},x^{+})-\Phi(z^{+},x^{-})-\Phi(z^{-},x^{+})+\Phi(z^{-},x^{-}))\\
 &  & -\Psi(z^{+},x^{+})+\Psi(z^{+},x^{-})-\Psi(z^{-},x^{+})+\Psi(z^{-},x^{-})\end{eqnarray*}
We expand these functions in $g^{2}$ as \[
P(q,Q)=P_{8}(q,Q)g^{8}+P_{10}(q,Q)g^{10}+P_{12}(q,Q)g^{12}+\dots\]
\[
\Sigma(q,Q)=1+\Sigma_{2}(q,Q)g^{2}+\Sigma_{4}(q,Q)g^{4}+\dots\]
The expansion of the rational part is quite straightforward and we
obtain \[
P_{8}(q,Q)=\frac{4096Q^{2}(-1+q^{2}+Q^{2})^{2}}{(q^{2}+Q^{2})^{4}(q^{4}+(-1+Q^{2})^{2}+2q^{2}(1+Q^{2}))}\]

\begin{eqnarray*}
\frac{P_{10}(q,Q)}{P_{8}(q,Q)} & = & -\frac{8(7q^{4}+3(Q^{2}-1)^{2}+10q^{2}(1+Q^{2}))}
{(q^{2}+Q^{2})(q^{4}+(-1+Q^{2})^{2}+2q^{2}(1+Q^{2}))}\end{eqnarray*}
\begin{eqnarray*}
\frac{P_{12}(q,Q)}{P_{10}(q,Q)} & = & -\frac{2(27+268q^{2}+704q^{4})}{(3+28q^{2}+64q^{4})
(q^{2}+Q^{2})}+\frac{16(-3q^{2}+2q^{2}Q)}{(1+4q^{2})(1+q^{2}-2Q+Q^{2})}\\
 &  & -\frac{16(3q^{2}+2q^{2}Q)}{(1+4q^{2})(1+q^{2}+2Q+Q^{2})}-\frac{8(-81q^{2}-117q^{4}+
 16q^{6}+27q^{2}Q^{2}+16q^{4}Q^{2})}{(3+16q^{2})(3+10q^{2}+7q^{4}-6Q^{2}+10q^{2}Q^{2}+3Q^{4})}\end{eqnarray*}
In expanding the $\Psi$ and $\Phi$ functions we use the same method
we used in \cite{Bajnok:2009vm}. The expansion of the $\Psi(x_{1},x_{2})$
functions for $\vert x_{2}\vert>1$ (string region) reads as follows
\begin{eqnarray*}
\Psi(x_{1},x_{2}) & = & -\frac{g}{x_{2}}(\Psi(1-ig(x_{1}+x_{1}^{-1}))+\Psi(1+ig(x_{1}+x_{1}^{-1})))\\
 &  & -\frac{ig^{2}}{2x_{2}^{2}}(\Psi_{1}(1-ig(x_{1}+x_{1}^{-1}))-\Psi_{1}(1+ig(x_{1}+x_{1}^{-1})))\\
 &  & +\frac{g^{3}}{2x_{2}}(\Psi_{2}(1-ig(x_{1}+x_{1}^{-1}))+\Psi_{2}(1+ig(x_{1}+x_{1}^{-1})))+\dots\end{eqnarray*}
where $\Psi_{n}(x)=(\frac{d}{dx})^{n}(\log(\Gamma(x))$ are the standard
polygamma functions. If $\vert x_{1}\vert>1$ then $\Phi(x_{1},x_{2})$
starts at $g^{6}$. In the opposite case using the identity $\Phi(x_{1},x_{2})=\Phi(0,x_{2})-\Phi(x_{1}^{-1},x_{2})$,
being valid if $\vert x_{1}\vert\neq1$, we can calculate the leading
expansion of $\Phi$ as 
\begin{equation}
\Phi(0,x)=\frac{2}{x}(\gamma_{E}g-3\zeta(3)g^{3}+\dots)
\end{equation}
Using functional identities valid for integer $Q$ we obtained\[
\Sigma_{2}(q,Q)=-\frac{16Q}{q^{2}+Q^{2}}-8\Bigr(S_{1}\Bigl(\frac{Q-iq-2}{2}\Bigl)+S_{1}\Bigl(\frac{Q+iq-2}{2}\Bigl)\Bigl)\]
and 
\begin{eqnarray}
\Sigma_{4}(q,Q) & = & \frac{1}{2}\Sigma_{2}(q,Q)^{2}+\frac{64Q(1+q^{2}+Q^{2})}{(q^{2}+Q^{2})^{2}}
-\frac{32iq}{q^{2}+Q^{2}}\Bigr(S_{2}\Bigl(\frac{Q-iq-2}{2}\Bigl)
-S_{2}\Bigl(\frac{Q+iq-2}{2}\Bigl)\Bigl)+\\
 &  & 32\Bigr(S_{1}\Bigl(\frac{Q-iq-2}{2}\Bigl)+S_{1}\Bigl(\frac{Q+iq-2}{2}\Bigl)\Bigl)
 +8\Bigr(S_{3}\Bigl(\frac{Q-iq-2}{2}\Bigl)+S_{3}\Bigl(\frac{Q+iq-2}{2}\Bigl)\Bigl)+18\zeta(3)
 \nonumber
 \end{eqnarray}
where $S_{n}(x)$ are the analytical continuation of the harmonic
sums $S_{n}(N)=\sum_{k=1}^{N}\frac{1}{k^{n}}$. They are related to
the polygamma functions as%
\footnote{For $n=0$ one has to replace $\zeta(1)$ with $\gamma_{E}$. %
} \[
\Psi_{n}(Q)=(-1)^{n+1}n!(\zeta(n+1)-S_{n+1}(Q-1)\]
Let us use now these expressions to calculate the four, five and six
loop wrapping corrections.

\section{Summary of the results}

Once we know the expansion of all functions we can systematically
calculate the wrapping corrections by evaluating the expression: \[
\Delta E=-\sum_{Q=1}^{\infty}\int\frac{dq}{2\pi}\left[g^{8}P_{8}+g^{10}(P_{10}+P_{8}
\Sigma_{2})+g^{12}(P_{12}+P_{10}\Sigma_{2}+P_{8}\Sigma_{4})+\dots\right]\]
Let us analyze the energy correction order by order.

\subsubsection*{4 loop contribution}

The first non-vanishing term is\[
\Delta E_{4}=-\sum_{Q=1}^{\infty}\int\frac{dq}{2\pi}P_{8}(q,Q)\]
The integrand can be decomposed as 

\[
P_{8}(q,Q)=\sum_{j=1}^{4}\frac{R_{8j}^{-}(Q)}{(q-iQ)^{j}}+\sum_{j=1}^{4}\frac{R_{8j}^{+}(Q)}{(q+iQ)^{j}}
+\frac{R_{8}^{--}(Q)}{q-iQ-i}+\frac{R_{8}^{+-}(Q)}{q+iQ-i}+\frac{R_{8}^{-+}(Q)}{q-iQ+i}+\frac{R_{8}^{++}(Q)}{q+iQ+i}\]
This form is actually quite generic, that is we can always expand
the rational part of the integrand as 

\[
P_{2k}(q,Q)=\sum_{\epsilon=\pm}\sum_{j=1}^{2k}\frac{R_{2k\, j}^{\epsilon}(Q)}{(q+\epsilon iQ)^{j}}
+\sum_{\epsilon_{1},\epsilon_{2}=\{\pm\}}\sum_{j=1}^{k}\frac{R_{2k\, j}^{\epsilon_{1}\epsilon_{2}}(Q)}{(q
+\epsilon_{1}iQ+\epsilon_{2}i)^{j}}\]
As the function $P_{8}$ is real $R_{8j}^{-}(Q)^{*}=R_{8j}^{+}(Q)$
and $R_{8}^{--}(Q)^{*}=R_{8}^{++}(Q)$, $R_{8}^{+-}(Q)^{*}=R_{8}^{-+}(Q)$.
The various coefficients can be expressed in negative powers of $Q$
and $2Q\pm1$ as \[
R_{84}^{+}(Q)=\frac{256}{Q^{2}}-\frac{512}{(2Q-1)}+\frac{512}{(2Q+1)}\qquad;
\qquad R_{83}^{+}(Q)=\frac{512i}{Q^{3}}-\frac{512i}{(2Q-1)^{2}}+\frac{512i}{(2Q+1)^{2}}\]
\[
R_{82}^{+}(Q)=-\frac{640}{Q^{4}}+\frac{512}{Q^{2}}+\frac{512}{(2Q-1)^{3}}-
\frac{512}{(2Q-1)}-\frac{512}{(2Q+1)^{3}}+\frac{512}{(2Q+1)}\]
\[
R_{81}^{+}(Q)=-\frac{640i}{Q^{5}}+\frac{512i}{Q^{3}}+\frac{512i}{(2Q-1)^{4}}-
\frac{512i}{(2Q-1)^{2}}-\frac{512i}{(2Q+1)^{4}}+\frac{512i}{(2Q+1)^{2}}\]
\[
R_{8}^{--}(Q)=-\frac{512i}{(2Q+1)^{4}}+\frac{512i}{(2Q+1)^{2}}\;;\quad R_{8}^{-+}(Q)=
\frac{512i}{(2Q-1)^{4}}-\frac{512i}{(2Q+1)^{2}}\]
We can further decompose any function as \[
R_{8j}^{\pm}=R_{8j,Q}^{\pm}+R_{8j,2Q+1}^{\pm}+R_{8j,2Q-1}^{\pm}\quad;\qquad R_{8j,Q}^{\pm}=
\sum_{k=1}^{5}\frac{R_{8j,Qk}^{\pm}}{Q^{k}}\quad;\quad R_{8j,(2Q\pm1)}^{\pm}=
\sum_{k=1}^{4}\frac{R_{8j,(2Q\pm1)k}^{\pm}}{(2Q\pm1)^{k}}\]
(It can be extended an analogous way for any $R_{2k\, j}$). We calculate
the integral by residues in which we take into account the poles on
the upper half plane. They are located at $q=iQ$ which we call the
kinematical pole and at $q=i(Q\pm1)$ which are the dynamical poles.
The contributions of the dynamical poles are canceled when summed
over the bound-state contributions ($Q$). Thus we have to keep only
the pole at $iQ$ which result in the following sum \[
\Delta E_{4}=-\sum_{Q=1}^{\infty}\mbox{Res}_{q=iQ}P_{8}(q,Q)=-
\sum_{Q=1}^{\infty}R_{81}^{-}(Q)=128(4\zeta(3)-5\zeta(5))\]
In summing up the $R_{81,2Q\pm1}$ terms here and later on we use
the following trick: \[
\sum_{Q=1}^{\infty}(R_{8j,2Q-1}^{\pm}(Q)+R_{8j,2Q+1}^{\pm}(Q))=
R_{8j,2Q-1}^{\pm}(1)+\sum_{Q=1}^{\infty}(R_{8j,2Q-1}^{\pm}(Q+1)+R_{8j,2Q+1}^{\pm}(Q))=0\]
This means that in the sum only the $Q^{-n}$ type terms survive,
which result in the sum $\sum_{Q=1}^{\infty}Q^{-n}$ and contains
single zeta functions reproducing the result of \cite{Bajnok:2008qj}.

\subsubsection*{5 loop contribution}

The second order correction has the form\[
\Delta E_{5}=-\sum_{Q=1}^{\infty}\int\frac{dq}{2\pi}(P_{10}(q,Q)+P_{8}(q,Q)\Sigma_{2}(q,Q))\]
Similarly to the 4 loop result in the first term only the residue
at $iQ$ contributes and we obtain a single sum: $P_{10}$ has a similar
decomposition as $P_{8}$ and only the $\sum_{Q=1}^{\infty}Q^{-n}=\zeta(n)$
terms survive which results in \[
-\sum_{Q=1}^{\infty}\int\frac{dq}{2\pi}P_{10}(q,Q)=-\sum_{Q=1}^{\infty}R_{10\,1,Q}^{-}(Q)=-128(48\zeta(3)+40\zeta(5)-105\zeta(7))\]
In calculating the integral of the term $P_{8}(q,Q)\Sigma_{2}(q,Q)$
we can use the fact that $P_{8}$ is real and replace in $\Sigma_{2}$
the expression $S_{1}(-1+\frac{Q+iq}{2})$ with its complex conjugate
$S_{1}(-1+\frac{Q-iq}{2})$: \[
\tilde{\Sigma}_{2}(q,Q)=-\frac{16Q}{q^{2}+Q^{2}}-16S_{1}\Bigl(\frac{Q-iq-2}{2}\Bigl)\]
 and take the real part of the integral. Interestingly the integral
is real in this case and we have\[
\int\frac{dq}{2\pi}(P_{8}(q,Q)\Sigma_{2}(q,Q))=\int\frac{dq}{2\pi}(P_{8}(q,Q)\tilde{\Sigma}_{2}(q,Q))=\Delta E_{5Q}+\Delta E_{5(2Q\pm1)}\]
 The replacement has the advantage that $S_{1}(-1+\frac{Q-iq}{2})$
does not have any pole on the upper half plane opposed to $S_{1}(-1+\frac{Q+iq}{2})$.
Thus calculating the integral we can take the residues at $iQ$ and
$i(Q\pm1)$ only. We use the same method in calculating the sums for
the $2Q\pm1$ type terms as we used for $\Delta_{4}$: we evaluate
the $Q=1$ term in the $(2Q-1)$ type terms, then we add the $Q\to Q+1$
shifted terms to the $2Q+1$ type terms. These terms include the dynamical
residues and the $R_{8j(2Q\pm1)}^{-}$ type kinematical residue and
result in \[
\Delta E_{5(2Q\pm1)}=\frac{256}{27}\pi^{4}(\pi^{2}-6)+2048\zeta(3)\]
. From the remaining $R_{8jQ}$ type terms at the kinematical residue
we got summands of the form $\sum_{Q=1}^{\infty}Q^{-n}S_{m}(Q-1)=\zeta(n,m)$,
for $n+m=4$ and $n+m=6$. Explicitly summing them up we obtained\[
\Delta E_{5Q}=512(8\zeta(2)^{2}-4\zeta(3)^{2}-11\zeta(2)\zeta(4)-8\zeta(2,2)+\zeta(2,4)-16\zeta(3,1)+4\zeta(3,3)+10\zeta(4,2)+20\zeta(5,1))\]
We can express the multiple zeta values in terms of products of simple
zeta functions either by the online high precision calculator of MZVs:
EZFACE, or analytically, by using the algebraic relations between them
\cite{Blumlein:1998if,Blumlein:2003gb} and the data mine for MZVs:
\cite{Blumlein:2009cf}. Combining all contributions we arrive at
the final 5 loop formula\[
\Delta E_{5}=-128(12\zeta(3)^{2}+32\zeta(3)+40\zeta(5)-105\zeta(7))\]

\subsubsection*{6 loop contribution}

The calculation of the six loop formula reduces to the evaluation
of \[
\Delta E_{6}=-\sum_{Q=1}^{\infty}\int\frac{dq}{2\pi}(P_{12}(q,Q)+P_{10}(q,Q)\Sigma_{2}(q,Q)+P_{8}(q,Q)\Sigma_{4}(q,Q))\]
The various terms have different structure in terms of multiple zeta
values. The first term can be evaluated just by taking the residue
at $iQ$ and summing up the $\sum_{Q=1}^{\infty}Q^{-n}=\zeta(n)$
type single zeta function contributions of $R_{12\,1,Q}^{-}(Q)$:
\[
128(576\zeta(3)+480\zeta(5)+392\zeta(7)-1701\zeta(9))\]
In evaluating the term $P_{10}\Sigma_{2}$ we use the trick to replace
$S_{1}(-1+\frac{Q+iq}{2})$ with its complex conjugate $S_{1}(-1+\frac{Q-iq}{2})$
and calculate the integral of $P_{10}\tilde{\Sigma}_{2}$ by taking
into account the contributions of the residues at $iQ$ and $i(Q\pm1)$
only. The results can be decomposed into the contributions of the
dynamical poles together with the $R_{10j(2Q\pm1)}^{-}$ type terms
of the kinematical pole and into the most complicated contribution
of the kinematical pole $R_{10jQ}^{-}$ terms. This latter one can
be written in terms of MZVs as $\sum_{Q=1}^{\infty}Q^{-n}S_{m}(Q-1)=\zeta(n,m)$,
for $n+m=4,6,8$. Explicitly summing up this term we obtain \begin{eqnarray*}
-1024(48\zeta(2)^{2}+(16-15\zeta(4))\zeta(4)+8\zeta(3)(\zeta(3)-6\zeta(5))+\zeta(2)(32\zeta(4)-105\zeta(6))-8\zeta(6)\\
-48\zeta(2,2)-80\zeta(3,1)+8\zeta(2,4)-8\zeta(3,3)-40\zeta(4,2)-88\zeta(5,1)\\
+15\zeta(4,4)+3\zeta(3,5)+45\zeta(5,3)+105\zeta(6,2)+210\zeta(7,1))\end{eqnarray*}
 The basis of the multiple zeta values contain an irreducible depth
two element which can be chosen to be $\zeta(3,5)$, see \cite{Blumlein:2009cf}.
Interestingly, however, using the relations between the MZVs, \cite{Blumlein:2009cf},
we could express the total contribution in terms of products of single
zeta values as\[
\frac{512}{675}(-1200\pi^{4}-80\pi^{6}+27\pi^{8}+60750\zeta(3)\zeta(5))\]
The other remaining terms give single zeta functions only. There are
two terms \begin{equation}
131072\frac{Q(1+Q)}{(1+2Q)^{4}}\Psi_{1}(Q)-65536\frac{Q(1+Q)}{(1+2Q)^{3}}\Psi_{2}(Q)-
\frac{4096}{(1+2Q)^{2}}\Psi_{3}(Q)+\frac{2048}{3(1+2Q)}\Psi_{4}(Q)\label{eq:sigma2rest}\end{equation}
and \begin{equation}
-131072\frac{Q(1+Q)}{(1+2Q)^{4}}\Psi_{1}(Q+\frac{1}{2})\label{eq:half1}\end{equation}
whose summation leads to alternating Euler sums. We do not write them
here as these terms will cancel when we combine them with other terms
coming from $P_{8}\Sigma_{4}$. 

In calculating $P_{8}\Sigma_{4}$, the trick of replacing $S_{n}(-1+\frac{Q+iq}{2})$
with its complex conjugate $S_{n}(-1+\frac{Q-iq}{2})$ and taking
the real part of the result can be applied. This works equally for
$S_{1}(-1+\frac{Q\pm iq}{2})S_{1}(-1+\frac{Q\pm iq}{2})$, but not
for $S_{1}(-1+\frac{Q+iq}{2})S_{1}(-1+\frac{Q-iq}{2})$ which is manifestly
real and needs a special care. For this reason, we separate $\Sigma_{4}$
as \[
\Sigma_{4}=\mbox{\ensuremath{\Re}e}\left(\tilde{\Sigma}_{4}\right)+64S_{1}(-1+\frac{Q+iq}{2})S_{1}(-1+\frac{Q-iq}{2})\]
where in $\tilde{\Sigma}_{4}$ the replacement can be applied and
it contains the rational part, too. The advantage of using $\tilde{\Sigma}_{4}$
is that it does not have any pole on the upper half plane, so in calculating
its contributions only the kinematical ($iQ$) an dynamical poles
($i(Q\pm1)$) have to be taken into account. In contrast, additionally
to these poles, $S_{1}(-1+\frac{Q+iq}{2})$ has simple poles at $q=i(2n+Q)$
with residue $2$ for any integers $n.$ Taking into account the additional
pole contributions we encounter a double sum of the form:

\[
\sum_{Q=1}^{\infty}\sum_{n=1}^{\infty}128\, P_{8}(i(2n+Q),Q)S_{1}(n+Q-1)\]
 We first separate the contributions coming from $R_{8j,Q}^{\pm}$
, $R_{8j,2Q\pm1}^{\pm}$ and $R_{8}^{\pm\pm},R_{8}^{\mp\pm}$ . In
calculating the $R_{8j,Q}^{\pm}$ part we use the result from the
Appendix to arrive at: \begin{eqnarray*}
2048(6\zeta(3)\zeta(4)+21\zeta(4)\zeta(2,1)-\zeta(3)\zeta(2,2)+8\zeta(2,3)-\zeta(2,5)-4\zeta(3)\zeta(3,1)+16\zeta(3,2)\\
-4\zeta(3,4)+24\zeta(4,1)+\zeta(2)(8\zeta(3)-19\zeta(5)-24\zeta(2,1)+\zeta(2,3)+4\zeta(3,2)+12\zeta(4,1))-10\zeta(4,3)\\
-20\zeta(5,2)-35\zeta(6,1)+16\zeta(2,2,1)-2\zeta(2,4,1)+32\zeta(3,1,1)-8\zeta(3,3,1)-20\zeta(4,2,1)-40\zeta(5,1,1)\end{eqnarray*}
In calculating the contributions of $R_{8j,2Q\pm1}^{\pm}$ we use
the usual trick to shift the $2Q-1$ type terms and combine them with
the $2Q+1$ type terms, but additionally we combine this step with
an appropriate shift $n\to n+1$, too. The resulting expression, apart
form a rational part, contains $\Psi(Q+1)$ with a quite long prefactor
which we do not write out explicitly here, as it will nicely combine
together with the contributions of the kinematical poles. The contributions
of $R_{8}^{\pm\pm},R_{8}^{\mp\pm}$ can be evaluated along the same
lines and we obtain \begin{equation}
\frac{262144Q(1+Q)}{(1+2Q)^{4}}\Psi(\frac{3}{2}+Q)\label{eq:half2}\end{equation}
Finally we have to calculate the contributions of the kinematical
and dynamical poles. The dynamical poles are simple and result in
poly gamma functions of the form $\Psi_{i}(\frac{1}{2}+Q)$ for $i=0,1$.
Combining their contributions with (\ref{eq:half1},\ref{eq:half2})
we obtain the simple result: \[
4096(28\zeta(3)-31\zeta(5))\]
The calculation of the contributions of the kinematical pole ($q=iQ$)
is the most complicated. It slightly simplifies when we collect the
remaining contributions from (\ref{eq:sigma2rest}) and from the term
coming from the $\Psi(Q+1)$ type term of the residues at $i(Q+2n)$.
The emerging sums, apart from rational expressions which can summed
up easily with Mathematica, contain the products of harmonic numbers
multiplied with rational functions. Decomposing them we have terms
of the form \begin{equation}
\sum_{Q=1}^{\infty}\frac{1}{Q^{k}}S_{n}(Q-1)S_{m}(Q-1)\label{eq:triplekinematical}\end{equation}
for $m+n=5,6,7$ or $8$ and alternating sums of the form \[
\sum_{Q=1}^{\infty}\left(\frac{1}{(-1+2Q)^{k}}-\frac{1}{(1+2Q)^{k}}\right)S_{n}(Q-1)S_{m}(Q-1)\]
 for $n,m$ and $k\leq5$. The latter one can be treated by the usual
trick of shifting $Q$ in the first term, and evaluating the $Q=1$
term separately. The difference of the resulting shifted harmonic
sums can be simplified into a single harmonic sum, which then can
be evaluated in terms of MZVs of depth two containing also individual
irreducible elements like $\zeta(3,5)$. In the expressions (\ref{eq:triplekinematical})
we can use that \[
S_{n}(Q-1)S_{m}(Q-1)=\sum_{j=1}^{Q-1}\frac{1}{j^{n}}\sum_{k=1}^{j-1}\frac{1}{k^{m}}+
\sum_{j=1}^{Q-1}\frac{1}{j^{m}}\sum_{k=1}^{j-1}\frac{1}{k^{n}}+S_{n+m}(Q-1)\]
This will lead to multivariate zeta functions of the form $\xi(k,n,m),\xi(k,m,n)$
and $\zeta(k,m+n)$. Unfortunately the emerging expressions are not
very illuminating and are rather lengthy. Magically, however, when
we combine all the contributions the irreducible $\zeta(3,5)$ term
disappears and we arrive at the nice result of the 6 loop correction
which is:

\begin{equation}
\Delta E_{6}=-128(48\zeta(3)^{2}-592\zeta(5)-24\zeta(3)(8+15\zeta(5))-322\zeta(7)+1701\zeta(9))
\label{eq:DE6}\end{equation}
In structure it is very similar to the result of other six loop calculations,
like the NLO result for the twist three \cite{Velizhanin:2010cm}
or single impurity \cite{Arutyunov:2010gu} operators. Let us finally
mention, that we evaluated numerically the integral form of $\Delta E_{6}$
and the result perfectly agreed with the exact analytical result (\ref{eq:DE6}).

\section{Conclusion}

In the paper we have evaluated Lüscher's finite size energy formula
for the simplest case in the AdS/CFT correspondence at NNLO and determined
the anomalous dimension of a single impurity operator in the $\beta=\frac{1}{2}$
deformed theory at six loops. The Lüscher formula expresses the energy
correction in terms of the scattering matrix of the physical particle
and the mirror bound-states. In calculating the weak coupling expansion
of the correction we have to integrate the expanded scattering matrix
elements for the mirror momentum and sum over the full bound-state
spectrum $Q\in\mathbb{Z}$. The expansion of the dressing phase in
the mirror-physical kinematics at order $n$ contains the $(n-1)^{th}$
power of the polygamma function. When integrated over the mirror momentum
by residues derivatives of polygamma functions evaluated at integer
values appear. These can be expressed in terms of nested harmonic
sums which after summation over the bound-state spectrum result in
multiple zeta values. At the $n^{th}$ order of the expansion MZVs
of depth $n$ appear, additionally, the degree of the MZV is increased
by 2 from order to order. Irreducible MZVs, those which cannot be
written in terms of products of zeta functions appear already at depth
2 and degree 8. Although during the calculation of the six loop anomalous
dimension of the single impurity operator irreducible MZVs appeared,
but in the final form they canceled out, leaving only products of
zeta functions. 

We have no explanation why such a cancellation appeared and it would
nice to evaluate the seven loop correction in this setting to see
whether a similar effect takes place at the next order, too. It would
be even more interesting to calculate the eight loop correction, since
there a new type of double wrapping effect will appear. Another interesting
problem is to evaluate these corrections using Laplace transformations
\cite{Bajnok:2009vm}, as in this case integrals of polylogarithms
appear. Then using the theory of motives just as for scattering amplitudes
in \cite{Goncharov:2010jf} might explain the form of the correction. 

The developed techniques can be extended to calculate the six loop
anomalous dimension of the Konishi operator, which corresponds to
a two particle state on the string side. We expect that the form of
the wrapping correction will be the same, only the actual coefficients
will change. This result would be very relevant to check to proposed
excited TBA equations \cite{Gromov:2009bc,Arutyunov:2009ax}. Indeed,
the comparison of the infinite coupled TBA equation to the Lüscher
correction at five loops \cite{Arutyunov:2010gb,Balog:2010xa,Arutyunov:2010gu}
confirmed only part of the TBA equations, while a six loop analysis
would check the full set.

The form of the correction we obtained might shed some light on how
a possible finite system of NLIE type integral equations would look
like as the weak coupling expansion of the anomalous dimension is
very similar in structure to the expansion of the BES equation \cite{Beisert:2006ez}.

\subsection*{Acknowledgments}

We thank Romuald Janik and Tomek Lukowski for the useful discussions,
for Matthias Staudacher to calling our attention to MZVs and for Johannes
Blümlein to reminding us the MZVs data mine. The work was supported
by a Bolyai Scholarship, and by OTKA K81461.

\appendix

\section{Summing up nested harmonic sums}

Here in this appendix we describe how we summed up nested harmonic
sums. 

At 6 loops we have triple and double sums and here we focus on the
triple sums only. Consider a sum of the form\[
f(a,b)=\sum_{Q=1}^{\infty}\frac{1}{Q^{a}}\sum_{n=1}^{\infty}\frac{1}{n^{b}}S_{1}(n+Q-1)\]
We claim that the result is \begin{eqnarray*}
f(a,b) & = & \zeta(a)\zeta(b,1)+\zeta(a)\zeta(b+1)\\
 &  & +\sum_{j=0}^{b-2}(-1)^{j}\zeta(b-j)\zeta(a,j+1)-(-1)^{b}\left(\zeta(a,b+1)+\zeta(a,b,1)\right)\end{eqnarray*}
where the Euler-Zagier sums are defined as \[
\zeta(a_{1},a_{2},\dots,a_{n})=\sum_{j_{1}>j_{2}>\dots>j_{n}>0}\frac{1}{j_{1}^{a_{1}}j_{2}^{a_{2}}\dots\, j_{n}^{a_{n}}}\]
The derivation goes as follows: 

First we use that $S_{1}(n+Q-1)=\sum_{j=1}^{n+Q-1}\frac{1}{j}$ and
decompose the sum as \[
\sum_{Q=1}^{\infty}\frac{1}{Q^{a}}\sum_{n=1}^{\infty}\frac{1}{n^{b}}\left(\sum_{j=1}^{n-1}\frac{1}{j}+\frac{1}{n}
+\sum_{j=1}^{Q-1}\frac{1}{n+j}\right)=\zeta(a)\zeta(b,1)+\zeta(a)\zeta(b+1)+
\sum_{Q=1}^{\infty}\frac{1}{Q^{a}}\left(\sum_{n=1}^{\infty}\frac{1}{n^{b}}
\sum_{j=1}^{Q-1}\frac{1}{n+j}\right)\]
Now exchanging the orders of the sums and using that\[
\sum_{j=1}^{Q-1}\left(\sum_{n=1}^{\infty}\frac{1}{n^{b}}\frac{1}{n+j}\right)=
\sum_{j=1}^{Q-1}\left(\sum_{k=0}^{b-2}\frac{(-1)^{k}}{j^{k+1}}\zeta(b-k)-
\frac{(-1)^{b}}{j^{b}}\left(\frac{1}{j}+\sum_{k=1}^{j-1}\frac{1}{k}\right)\right)\]
we can write\[
\sum_{Q=1}^{\infty}\frac{1}{Q^{a}}\left(\sum_{n=1}^{\infty}\frac{1}{n^{b}}
\sum_{j=1}^{Q-1}\frac{1}{n+j}\right)=\sum_{k=0}^{b-2}(-1)^{k}\zeta(a,k+1)
\zeta(b-k)-(-1)^{b}\left(\zeta(a)\zeta(b+1)+\zeta(a,b,1)\right)\]

An easier manipulation gives \begin{eqnarray*}
\sum_{Q=1}^{\infty}\frac{1}{Q^{a}}\sum_{n=1}^{\infty}\frac{1}{(n+Q)^{b}}S_{1}(n+Q-1) &
= & \zeta(a)\zeta(b,1)-\zeta(a+b,1)-\zeta(a,b,1)\end{eqnarray*}
In the 6 loop calculation terms with $b=1$ appear. They are individually
divergent but their appearing combination is finite.

\end{document}